\documentclass[twocolumn,tighten]{aastex701}

\usepackage{amsmath,amssymb,framed}
\usepackage{bm}
\usepackage{balance}

\def\bq{\begin{equation}}
\def\eq{\end{equation}}
\def\bqy{\begin{eqnarray}}
\def\eqy{\end{eqnarray}}

\def\p{\partial}

\def\p{\partial}

\def\R{\mathbb{R}}

 \def\q#1#2{q^{\hspace{1pt}#1}_{\,,\hspace{1pt}#2}}
 \def\a#1#2{a^{\hspace{1pt}#1}_{\,,\hspace{1pt}#2}}
 \def\d#1#2{\delta^{\hspace{1pt}#1}_{\,,\hspace{1pt}#2}}

\begin{document}
\title{\large{Transition of Magnetic Reconnection Regimes in Partially Ionized Plasmas}}

\author[0000-0003-4821-3120]{Liang Wang}
\affiliation{Center for Space Physics and Department of Astronomy, Boston University, Boston, MA 02215, USA}
\email[show]{wang0734@bu.edu}

\author[0000-0002-8990-094X]{Chuanfei Dong}
\affiliation{Center for Space Physics and Department of Astronomy, Boston University, Boston, MA 02215, USA}
\affiliation{School of Natural Sciences, Institute for Advanced Study, Princeton, NJ 08540, USA}
\email[show]{dcfy@bu.edu}

\author[0000-0002-4237-2211]{Yi-Min Huang}
\affiliation{Department of Astronomy, University of Maryland, College Park, Maryland 20742, USA}
\affiliation{National Aeronautics and Space Administration Goddard Space Flight Center, Greenbelt, MD 20771, USA}
\email{yopology@umd.edu}

\author[0009-0002-1692-8391]{Yue Yuan}
\affiliation{Center for Space Physics and Department of Astronomy, Boston University, Boston, MA 02215, USA}
\affiliation{Department of Earth, Planetary, and Space Sciences, University of California, Los Angeles, Los Angeles, CA, 90095, USA}
\email{yueyuan2003@g.ucla.edu}

\author[0000-0003-1553-6337]{Xinmin Li}
\affiliation{Center for Space Physics and Department of Astronomy, Boston University, Boston, MA 02215, USA}
\email{xli8@bu.edu}

\author[0000-0002-4168-9225]{Yang Zhang}
\affiliation{Department of Astrophysical Sciences, Princeton University, Princeton, NJ 08544, USA}
\affiliation{University Corporation for Atmospheric Research, Boulder, CO 80301, USA}
\email{yz0172@princeton.edu}

\begin{abstract}

Magnetic reconnection in partially ionized plasmas plays a crucial
role in a wide range of solar, astrophysical, and laboratory environments.
While reconnection in such plasmas is commonly characterized by the ion-neutral
coupling strength and the ionization fraction $\chi=n_{i}/(n_{i}+n_{n})$, most previous studies have focused primarily on the former.
A systematic exploration of the ionization fraction, particularly
in combination with ion-neutral coupling, is still lacking. This study
presents the first systematic scan of the two-dimensional parameter
space defined by ion-neutral collisionality and ionization fraction,
enabling investigation of the transition from strongly coupled reconnection
to faster, decoupled reconnection. To achieve this, we employ a new
three-fluid, five-moment numerical model that treats electrons, ions,
and neutrals as separate species on an equal footing. We find that
in the strongly coupled regime, the reconnection rate is consistent
with a $\chi^{1/4}$ scaling. As collisionality decreases, the system
transitions to a fast, ionization-independent regime. On the other
hand, in the weakly coupled and fast-reconnection regimes, the current
sheet approaches an ion-inertial-scale thickness rather than the expanded
hybrid scale $d_{i}\chi^{-1/2}$ predicted by fully coupled analytic
fluid theories. The identified critical thickness
and the resulting onset of fast reconnection agree reasonably well
with recent fully kinetic simulations and laboratory experiments.
In addition, we show that, over a wide range of coupling strengths,
the ion outflow velocities remain Alfvénic, scaling with the appropriate
ion or hybrid Alfvén speed, while the hybrid outflow velocity scales
as $\chi^{1/2}$ when normalized by ion Alfvén speed.
\end{abstract}

\section{\label{sec:introduction}Introduction}

Magnetic reconnection is a fundamental plasma process characterized by the breaking and topological reconnection of magnetic field lines, facilitating the rapid conversion of magnetic energy into kinetic and thermal energy~\citep{reconnection1,reconnection2,Ji2022}. While historically studied in fully ionized regimes, reconnection is common in partially ionized plasmas, such as the solar chromosphere \citep{Shibata2007, Tian2014, reconnection4}, protostellar disks \citep{Oberg2011}, and the interstellar medium \citep{Fielding2023}. In these environments, interactions between charged particles and neutrals can strongly influence the structure of the current sheet and modify both the onset criteria and the rate of energy dissipation. {Ideal-tearing models have also been applied to partially ionized layers of the solar atmosphere and protoplanetary disks, where successive current-sheet disruptions may connect coupled large scales to decoupled fast reconnection at small scales~\citep{Pucci2024}.} {As a recent example, two-fluid ion-neutral magnetohydrodynamic (MHD) studies of plasmoid coalescence confirmed that adding neutral dynamics, and later ionization/recombination, can modify reconnection dynamics relative to fully ionized single-fluid MHD models~\citep{Murtas2021,Murtas2022}.}

Theories for reconnection in partially ionized plasmas are often defined by the coupling regimes between ions and neutrals: weak, intermediate, or strong coupling~\citep{Zweibel1989,Malyshkin2011}. {Linear tearing theory provides a complementary onset perspective: ion-neutral collisions modify both the tearing growth rate and the critical current-sheet aspect ratio for fast disruption across these coupling regimes~\citep{Pucci2020}.} In the strongly coupled limit, where the ion-neutral collision frequency $\nu_{in}$ is high relative to the transit time through the current sheet, ions and neutrals move together, effectively increasing the bulk inertia by a factor of  $\chi$, defined as
the ionization fraction $\chi=n_{i}/(n_{i}+n_{n})$, where $n_i$ and $n_n$ are the ion and neutral densities, respectively. One immediate consequence is the slowdown of the bulk Alfvén speed, and thus the Lundquist number is reduced, by a factor of ${\chi}^{1/2}$. The steady-state resistive Sweet-Parker reconnection rate, normalized over the ion Alfvén speed, is correspondingly reduced by a factor of $\chi^{1/4}$, while the Sweet-Parker current sheet thickness is thickened to $\delta_{\rm SP}^*=\chi^{-1/4} \delta_{\rm SP}$.  On the other hand, the enhanced inertia also leads to the broadening of the inertial length $d_i$ by a factor of $\chi^{-1/2}$ to $d^*=\chi^{-1/2}d_i$. It was predicted that reconnection would transition to a fast regime once the Sweet–Parker current sheet thins to the hybrid inertial length.

Recent studies showed interesting discrepancies between predictions from fully coupled fluid models and results from kinetic approaches. Laboratory experiments at MRX \citep{Lawrence2013} and fully kinetic Particle-In-Cell (PIC) simulations \citep{JJAPRL} indicate that the current sheet thickness does not expand to the hybrid inertial length $d^*$ but rather stays at ion inertial length $d_i$, regardless of the ionization fraction.
The comparative study of \citet{JJAPOP} further found that multi-fluid MHD simulations tend to maintain a coupled state and produce a slower, Sweet-Parker-like scaling.
In contrast, at higher ionization fraction ($\chi>0.15$), fully kinetic simulations allow a transition to faster, decoupled reconnection, where the reconnection rate scales like $\chi^{1/2}$ when normalized using the Alfvén speed based on the ion mass density $\rho_i$, indicating fast reconnection rate at the bulk Alfvén speed.

The discrepancies noted above point to a gap between fluid and kinetic descriptions. In this study, we address this gap using a five-moment, three-fluid model that treats electrons, ions, and neutrals as separate species. We scan ion–neutral collisionality and ionization fraction to identify the transition from strongly coupled, resistive reconnection to decoupled, Hall-mediated reconnection. {In this sense, the present study complements two-fluid ion-neutral MHD comparisons with fully ionized single-fluid MHD models~\citep{Murtas2021,Murtas2022} by moving one step further in the model hierarchy toward separately evolved electron, ion, and neutral fluids.} We then compare the resulting reconnection behavior with earlier kinetic and multifluid MHD results to evaluate whether the five-moment model can capture the onset, rate, and current-sheet structure of fast reconnection seen in kinetic regimes. We do not include charge exchange, ionization, or recombination, as our goal is to isolate the role of frictional coupling between ions and neutrals.

The paper is structured as follows: Sec.~\ref{sec:numerical} details the numerical model and the simulation setup employed in this study; Sec.~\ref{sec:results} presents the primary simulation results; in Sec.~\ref{sec:discussion}, we further discuss the findings and their broader implications.

\section{\label{sec:numerical}Numerical Model and Simulation Setup}

The Gkeyll five-moment multifluid model~\citep{hakim2006high,wang2020exact} is used in this study. This model evolves the density, momentum, and energy equations of electron, ion, and neutral species (denoted by subscripts $e$, $i$, and $n$) separately. The number density equations take the form 
\begin{equation}
	\frac{\partial\left(m_{s}n_{s}\right)}{\partial t}+\nabla\cdot\left(m_{s}n_{s}\mathbf{u}_{s}\right)=0,\quad s=e,i,n,
\end{equation}
where $m_{s}$, $n_{s}$, and $\mathbf{u}_{s}$ are the mass, number density, and bulk velocity of species $s$. The momentum equations
are 
\begin{align}
	\frac{\partial\left(m_{e}n_{e}\mathbf{u}_{e}\right)}{\partial t}+\nabla\cdot\left(m_{e}n_{e}\mathbf{u}_{e}\mathbf{u}_{e}\right)+\nabla p_{e} \\ \nonumber
	=n_{e}q_{e}\left(\mathbf{E}+\mathbf{u}_{e}\times\mathbf{B}\right)+\mathbf{R}_{ei},\\
	\frac{\partial\left(m_{i}n_{i}\mathbf{u}_{i}\right)}{\partial t}+\nabla\cdot\left(m_{i}n_{i}\mathbf{u}_{i}\mathbf{u}_{i}\right)+\nabla p_{i} \\ \nonumber
	=n_{i}q_{i}\left(\mathbf{E}+\mathbf{u}_{i}\times\mathbf{B}\right)+\mathbf{R}_{in}+\mathbf{R}_{ie},\\
	\frac{\partial\left(m_{n}n_{n}\mathbf{u}_{n}\right)}{\partial t}+\nabla\cdot\left(m_{n}n_{n}\mathbf{u}_{n}\mathbf{u}_{n}\right)+\nabla p_{n} & =\mathbf{R}_{ni}. 
\end{align}
where $q_{s}$ is the charge of species $s$. In particular, $q_{n}=0$, $q_{i}=-q_{e}=e$.

The ion-neutral and neutral-ion collision terms are formulated to conserve total momentum between the two species, 
\begin{align}
	\mathbf{R}_{in}=\nu_{in}n_{i}m_{i}\left(\mathbf{u}_{n}-\mathbf{u}_{i}\right),      \\
	\quad\mathbf{R}_{ni}=\nu_{ni}n_{n}m_{n}\left(\mathbf{u}_{i}-\mathbf{u}_{n}\right), 
\end{align}
where the collision frequency is calculated during the
simulation by
\begin{align}
	\nu_{in}=\Sigma_{in}\sqrt{\frac{8k_{B}T_{in}}{\pi m_{in}}}\frac{n_{n}m_{n}}{m_{n}+m_{i}},      \\
	\quad\nu_{ni}=\Sigma_{in}\sqrt{\frac{8k_{B}T_{in}}{\pi m_{in}}}\frac{n_{i}m_{i}}{m_{n}+m_{i}}, 
\end{align}
with $\Sigma_{in}=\Sigma_{ni}$ being the collisional cross-section
of choice, $m_{ni}=m_{in}\equiv m_{i}m_{n}/\left(m_{i}+m_{n}\right)$
being the reduced mass, and $T_{in}=T_{ni}\equiv\left(m_{i}T_{n}+m_{n}T_{i}\right)/\left(m_{i}+m_{n}\right)$
being the reduced temperature. For this study, $\Sigma_{in}$
	is a constant. {This prescription uses the thermal relative speed in
	an elastic hard-sphere collision frequency. It does not include additional
	drift-dependent ion-neutral processes, such as charge exchange, which can
	provide momentum and energy exchange when $|\mathbf{u}_i-\mathbf{u}_n|$
	becomes comparable to or larger than the ion-neutral thermal
	speed~\citep[e.g.,][]{Murtas2021,Murtas2022}. Such
	effects could shift the effective coupling boundary and modify the detailed
	scaling behavior, although the overall transition from strongly coupled to
	decoupled reconnection is expected to remain. Quantifying such
	drift-dependent cross sections is left for future work.}

{The electron-ion friction terms use the same conservative momentum-exchange
form, with $\mathbf{R}_{ei}=-\mathbf{R}_{ie}$, but the local collision frequency
is specified in a Coulomb-like scaling,
\begin{equation}
	\nu_{ei}=\nu_{ei,0}\frac{n_e}{n_0}\left(\frac{T_{e0}}{T_e}\right)^{3/2},
\end{equation}
where $\nu_{ei,0}$ is the reference collision frequency at the initial uniform density and temperature.
In all simulations, the reference collision
frequency $\nu_{ei,0}$, or the corresponding effective resistivity, is inferred
from a fixed effective Lundquist number $S_{\rm reference}=L_x v_{A0}/\eta={10}^5$ using the uniform initial
temperature.}

Finally, the energy equations take the
form
\begin{align}
	\frac{\partial\mathcal{E}_{e}}{\partial t}+\nabla\cdot\left[\mathbf{u}_{e}\left(p_{e}+\mathcal{E}_{e}\right)\right] & =n_{e}q_{e}\mathbf{u}_{e}\cdot\mathbf{E}+\mathbf{u}_{e}\cdot\mathbf{R}_{ei}+Q_{ei},        \\
	\frac{\partial\mathcal{E}_{i}}{\partial t}+\nabla\cdot\left[\mathbf{u}_{i}\left(p_{i}+\mathcal{E}_{i}\right)\right] & =n_{i}q_{i}\mathbf{u}_{i}\cdot\mathbf{E}+\mathbf{u}_{i}\cdot\mathbf{R}_{in} \notag \\
	                                                                                                                    & \quad+\mathbf{u}_{i}\cdot\mathbf{R}_{ie}+Q_{in}+Q_{ie}, \\
	\frac{\partial\mathcal{E}_{n}}{\partial t}+\nabla\cdot\left[\mathbf{u}_{n}\left(p_{n}+\mathcal{E}_{n}\right)\right] & =\mathbf{u}_{n}\cdot\mathbf{R}_{ni}+Q_{ni},                                         
\end{align}
where $\mathcal{E}_{s}=\frac{p_{s}}{\gamma-1}+\frac{1}{2}m_{s}n_{s}\left|\mathbf{u}_{s}\right|^{2}$
is the total (internal plus kinetic) energy, $\gamma=5/3$, and $p_{s}$
is the species' scalar pressure. {The terms $\mathbf{u}_{s}\cdot\mathbf{R}_{sr}$
are the mechanical work done by collisional drag on the bulk flow, while
$Q_{sr}$ denotes the frictional heating.} For ion-neutral
collisions, these heating terms are 
\begin{align}
	Q_{in} & =\frac{n_{i}m_{i}\nu_{in}}{m_{i}+m_{n}}\left[3\left(\frac{p_{n}}{n_{n}}-\frac{p_{i}}{n_{i}}\right)+m_{n}\left|\mathbf{u}_{n}-\mathbf{u}_{i}\right|^{2}\right],  \\
	Q_{ni} & =\frac{n_{n}m_{n}\nu_{ni}}{m_{i}+m_{n}}\left[3\left(\frac{p_{i}}{n_{i}}-\frac{p_{n}}{n_{n}}\right)+m_{i}\left|\mathbf{u}_{n}-\mathbf{u}_{i}\right|^{2}\right],
\end{align}
where the adopted gas gamma $\gamma=5/3$ has been applied.
{The electron-ion
collisional energy exchange is implemented with the same split between
mechanical work and internal-energy heating, using the Coulomb-like electron-ion
collision frequency defined above. More detailed variants of the fluid collision
equations can be found in \citet{PopescuBraileanu2019,schunk2009,Meier2012}.}
{Electron-neutral collisions are neglected as they are not expected to play a leading role in reconnection dynamics in the parameter regime investigated here. They may, however, contribute to Ohmic diffusion and localized heating in partially ionized regions of the chromosphere, and their quantitative impact will be assessed in future work.}

The electromagnetic field is evolved by the full Maxwell's equations
\begin{align}
	\frac{1}{c^{2}}\frac{\partial\mathbf{E}}{\partial t}+\mu_{0}\mathbf{J} & =\nabla\times\mathbf{B},  \\
	\frac{\partial\mathbf{B}}{\partial t}                                  & =-\nabla\times\mathbf{E}, 
\end{align}
akin to kinetic approaches. The fluid species and the electromagnetic field are coupled
through the source terms on the right-hand-side of the momentum and energy equations.
Resonant charge exchange between the ion and neutral populations is not included in this study. To assess its potential impact, we estimated the momentum transfer rates from both elastic collision and charge exchange using the charge exchange formulation from the multifluid MHD model in \citet{JJAPOP}. The two rates were found to be of the same order of magnitude. 
Thus, we argue that within the parameter regime investigated in this study, neglecting charge exchange is unlikely to alter the fundamental reconnection physics or scaling laws. 

In this investigation, we set up the five-moment simulations to resemble
those used by \citet{JJAPOP} but with some modifications. We work in the $x$-$y$ domain, with $\hat{x}$ and $\hat{y}$ being the inflow and outflow direction of the magnetic reconnection, $\hat{z}$ being the out-of-plane direction. The neutral species was initialized as a uniform background at rest. An ionization fraction is defined by
\begin{equation}
	\chi_0=\frac{n_{b}}{n_{n}+n_{b}},
\end{equation}
where the number densities $n_{n}$ and $n_{b}$ correspond to the uniform
neutral background and the ionized plasma, respectively (see below). The unperturbed
plasma forms the Harris current sheet of thickness $\delta$, with
number density and magnetic field,
\begin{equation}
	n_{i}\left(x,y\right)=n_{e}\left(x,y\right)=n_{b}+n_{0}\mathrm{sech}^{2}\left(\frac{y}{\delta}\right),
\end{equation}
\begin{equation}
	\mathbf{B}\left(x,y\right)=\hat{z}B_{0}\tanh\left(\frac{y}{\delta}\right).
\end{equation}
The total electric current, 

\begin{equation}
	\mathbf{J}\left(x,y\right)=\frac{1}{\mu_{0}}\nabla\times\mathbf{B}_{0}\left(x,y\right)=-\hat{z}\frac{B_{0}}{\mu_{0}\delta}\frac{1}{\cosh^{2}\left(y/\delta\right)}
\end{equation}
contains contributions from both electron and ion species due to the
diamagnetic drift, so that $J_{ze0}/J_{zi0}=T_{e0}/T_{i0}$, where
$T_{e0}$ and $T_{i0}$ are the uniform electron and ion temperature.
In this investigation, $T_{e0}=T_{i0}=T_{0}$, though white noises are added to the electron temperature to break the initial configuration symmetry. A long-wavelength perturbation
is imposed on the magnetic field, $\delta\mathbf{B}=\hat{z}\times\nabla\delta\psi$,
to introduce a single X-line initially, 
\begin{equation}
	\delta\psi\left(x,y\right)=\delta\psi_{0}\cos\left(2\pi x/L_{x}\right)\cos\left(\pi y/L_{y}\right).
\end{equation}
Here, $d_{i0}=\sqrt{m_{i}/\left(n_{0}q^{2}\mu_{0}\right)}$ denotes the ion inertial length associated with $n_{0}$. The characteristic parameters include $m_{n}=m_{i}+m_{e}$, $T_{n0}=T_{i0}=T_{e0}=T_{0}$,
and
\[
	\frac{m_{i}}{m_{e}}=25,\,~\frac{\text{\ensuremath{\omega_{pe0}}}}{\Omega_{ce0}}=2,\,~\beta_{0}=\frac{2\mu_{0}n_{b}T_{0}}{B_{0}^{2}}=0.3,
\]
\begin{equation}
	n_b/n_0=0.3,~\frac{\delta}{d_{i0}}=4,\,~\frac{\delta\psi_{0}}{B_{0}d_{i0}}=0.1.
\end{equation}
Here, a thick initial current sheet is used to fully capture the thinning process.

The reference ion mean-free-path $\lambda_{\rm mfp,in}\equiv\left(n_{n}\Sigma_{in}\right)^{-1}$
is used to determine the constant collisional cross-section  $\Sigma_{in}$ indirectly.
{In this investigation, the reference electron-ion collision frequency
$\nu_{ei,0}$ is inferred from a fixed effective Lundquist number $S=10^5$
based on the uniform initial temperature.} We then vary
\begin{equation*}
    \frac{\lambda_{\rm mfp,in}}{d_{i0}}=0.05,~0.5,~5,~50,
\end{equation*}
to cover strong to weakly ion-neutral coupling, and simultaneously varying initial ionization fraction
\begin{equation*}
    \chi_0=0.02,~0.04,~0.1,~0.2,~0.4,~0.8,
\end{equation*}
to span weakly to highly ionization regimes.
A total of 24 runs were performed to cover the parameter matrix.
The domain extends $L_{x}\times L_{y}=400d_{i0}\times200d_{i0}$ and
is discretized on a $4096\times2048$ grid, with periodic and conducting
wall boundary conditions at $\hat{x}$ and $\hat{y}$ boundaries,
respectively. The above dimensionless parameters characterized the systems, though we further adopted unit normalizations for the numerical implementation such that $\varepsilon_0=\mu_0=m_i=q_i=n_0=1$.

\section{\label{sec:results}Simulation Results}

The top row of Figure 1 gives an overview of the reconnection rate evolution.
In the strongly collisional limit ($\lambda_{\rm mfp,in}/d_{i0}=0.05$, top left panel), there is a clear stratification in reconnection rates based on ionization, with higher $\chi_0$ leading to earlier onset times and higher peak reconnection rates.
Conversely, as the ion-neutral collisionality decreases ($\lambda_{\rm mfp,in}/d_{i0}=50$, top right panel), the system approaches a decoupled regime so that the temporal evolution and peak magnitude of the reconnection rates become largely insensitive to the ionization fraction.
{The intermediate case \(\lambda_{\rm mfp,in}/d_{i0}=5\) shows a mild exception to the otherwise smoother ionization trend: the \(\chi_0=0.02\) and \(0.04\) runs reach slightly higher peak rates than the higher-ionization cases. This reversal is weak, and should not be interpreted as a separate scaling trend. Instead, once secondary plasmoids are present, their growth and merger can enhance the instantaneous reconnection rate in an impulsive manner, introducing uncertainty in the exact peak value achieved in a given run.}
The onset time for fast reconnection in this five-moment model ranges from approximately $1500$ to $4700\Omega_{ci0}^{-1}$. These timescales are comparable to those reported in previous multi-fluid MHD simulations but are significantly longer than the characteristic onset times of $100$-$200\Omega_{ci0}^{-1}$ typically observed in fully kinetic simulations by \citet{JJAPOP,JJAPRL}.
In addition to the inherent differences in onset between fluid and kinetic regimes, this discrepancy is likely largely attributable to the kinetic simulations being initialized with a thin current sheet with a half-thickness of $d_{i0}$.

The row 2-4 of Figure 1 shows the current layer structures from selected runs.
In this strongly neutral-dominated regime ($\chi_0=0.02$, second row), decreasing the collisionality (moving left to right) results in an increase in maximum outflow speeds and a transition in the current sheet structure, consistent with a transition from coupled to decoupled dynamics.
At intermediate ionization ($\chi_0=0.1$, third row), structural changes are less significant. In the least collisional case (right panel), the early formation of secondary plasmoids becomes visible. In this three-fluid five-moment numerical investigation, dynamic plasmoid formation is generally more pronounced in low-collisionality or highly ionized runs during their nonlinear stages.
For the highest ionization ($\chi_0=0.8$, bottom row), the outflow structure remains largely invariant across the range of collisionalities. This indicates that at high ionization fractions, the influence of ion-neutral collisions is negligible, and the system behaves similarly to the fully ionized limit.

\begin{figure*}
\includegraphics[width=1\textwidth]{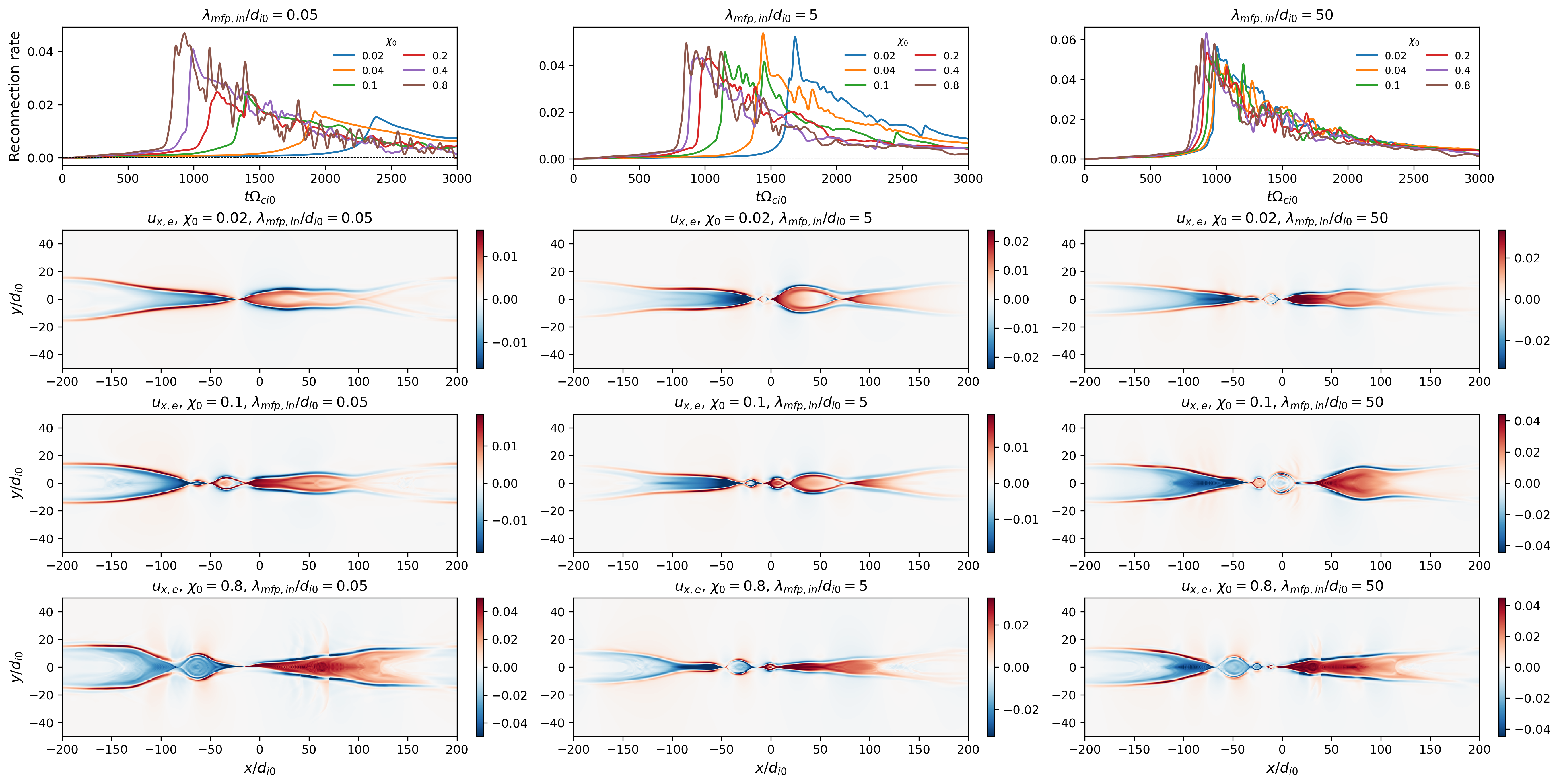}

\caption{Overview of the temporal evolution of reconnection rates and the
spatial structure of electron outflows across the parameter space.
(Top Row): Time-dependent reconnection rates normalized by the product
of the asymptotic reconnecting magnetic field, $B_{0}$, and the ion
Alfvén speed, $v_{A0}=B_{0}/\sqrt{\mu_{0}m_{i}n_{b}}$. The rates are computed as the time derivative of the globally reconnected flux, i.e., $\max(\psi)-\min(\psi))$ along the midplane, $\psi$ is the magnetic flux function. The columns
correspond to runs with increasing ion-neutral collisionality, parameterized
by the ratio of the mean free path to the ion inertial length, $\lambda_{{\rm mfp,in}}/d_{i0}$.
Within each panel, curves represent different initial ionization fractions,
$\chi_{0}$. (Rows 2--4): 2D snapshots of electron outflow velocities (normalized by the speed of light)
taken at the time of the first primary peak in the reconnection rate.
Each row corresponds to a specific ionization fraction. Note that
such primary peak times may precede the global maximum rate in cases
where secondary plasmoid formation occurs later.}\label{fig:overview}

\end{figure*}

The scaling of peak reconnection rate versus ionization fraction or ion-neutral collisionality are drawn in Figure 2.
{Figure 2a shows the ionization dependence. At high collisionality (low $\lambda_{\rm mfp,in}/d_{i0}$), the reconnection rate follows the qualitative trend expected from the fully coupled Sweet-Parker estimate. Because the available ionization range spans less than two decades and the scatter is non-negligible, the $\chi_0^{1/4}$ line in Figure~2a should be interpreted as a theoretical reference rather than a fitted or independently validated scaling.} {The slight enhancement of the \(\chi_0=0.02\) and \(0.04\) rates in the \(\lambda_{\rm mfp,in}/d_{i0}=5\) set reflects the same behavior visible in the corresponding Figure~\ref{fig:overview} rate traces, and should be viewed as quantitative scatter associated with plasmoid-mediated variability rather than a separate scaling trend.} As collisionality decreases, the rate transitions to a fast, ionization-independent regime characteristic of Hall-mediated reconnection.

Figure 2b examines the collisionality dependence. Generally, weaker ion-neutral coupling (higher $\lambda_{\rm mfp,in}/d_{i0}$) enhances the reconnection rate. The $\lambda_{\rm mfp,in}^{1/4}$ scaling observed up to $\lambda_{\rm mfp,in}/d_{i0} \approx 5$ suggests an intermediate regime where ion-neutral friction plays a governing role. In the high ionization limit ($\chi_0=0.8$), the rate becomes insensitive to collisionality, consistent with the fully ionized limit. 

Next, we move to better understand the nature of onset, Figure 2c summarizes the critical current sheet half-thickness across various runs. {For the strongest coupling shown, $\lambda_{\rm mfp,in}/d_{i0}=0.05$, the onset thickness decreases with increasing $\chi_0$, but the dependence is weaker than the fully coupled $\chi_0^{-1/4}$ scaling. For weaker coupling, $\lambda_{\rm mfp,in}/d_{i0}=5$ and $50$, the onset thickness is nearly independent of $\chi_0$ and remains close to $0.7\,d_{i0}$. The exact numerical value of this thickness should not be over-interpreted, because the onset time cannot be determined in a fully universal way and can be affected by plasmoid formation and other details of the reconnection evolution. Nevertheless, across most runs the onset thickness remains in the relatively narrow range $0.7$--$1\,d_{i0}$, broadly consistent with the ion-inertial-scale onset found in kinetic simulations by \citet{JJAPOP}.} {Thus, the present scan supports a transition picture rather than a precision measurement of the asymptotic coupled-fluid exponent: weakly coupled runs show a nearly ion-inertial-scale onset thickness, while stronger coupling introduces a progressively clearer $\chi_0$-dependence. Additional run sets at $\lambda_{\rm mfp,in}/d_{i0}<0.05$ would be required to determine whether this trend continues to steepen toward the fully coupled $\chi_0^{-1/4}$ prediction.}

\begin{figure*}
\includegraphics[width=1\textwidth]{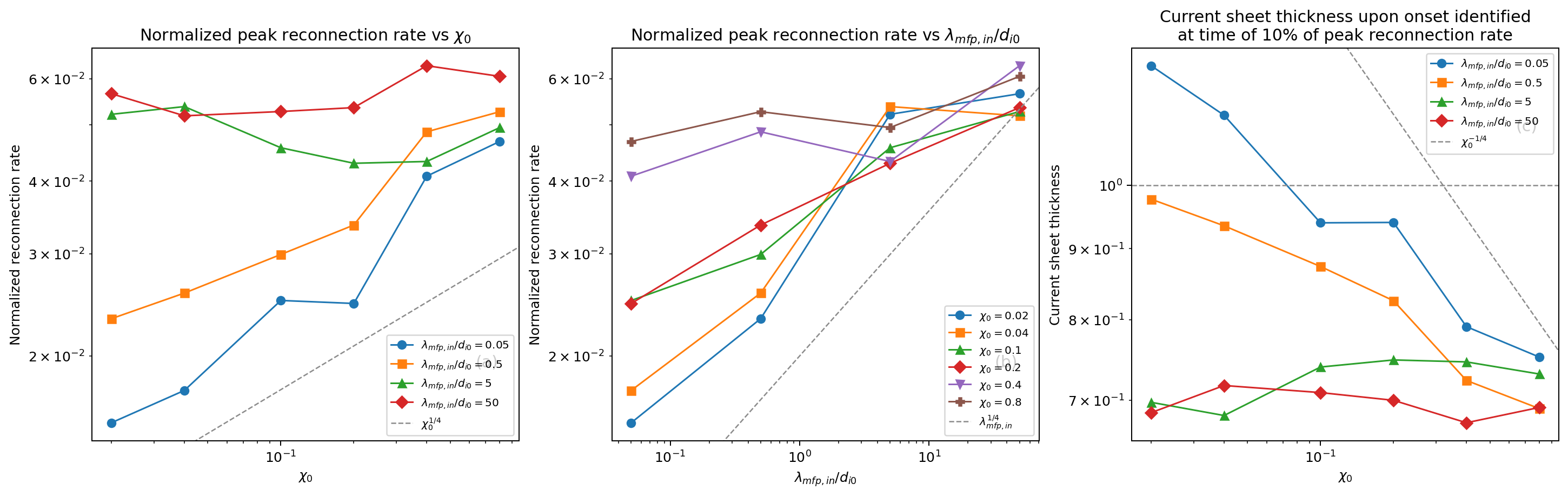}

\caption{Scaling dependencies of the reconnection rate and current sheet geometry
at onset. (a) Peak reconnection rate as a function of ionization fraction,
$\chi_{0}$. Data points on the same curves represent the same ion-neutral
collisionality, $\lambda_{{\rm mfp,in}}/d_{i0}$. The dashed line
indicates the theoretical $\chi^{1/4}$ scaling for the resistive
Sweet-Parker regime. (b) Peak reconnection rate as a function of the
collisionality parameter, $\lambda_{{\rm mfp,in}}/d_{i0}$, for fixed
$\chi_{0}$. A $\lambda_{{\rm mfp,in}}^{1/4}$
reference scaling is provided. (c) Scaling of the critical current
sheet half-thickness, $\delta_{\rm CS}$, measured at the onset time (defined
as 10\% of the peak reconnection rate). The dashed slanted line represents
the theoretical prediction for fully coupled fluids, while the horizontal
dashed line marks the ion inertial length, $d_{i0}$.}\label{fig:rate_and_onset_thickness_scaling}
\end{figure*}

Figure 3 illustrates the geometric properties of the current sheet during the fast reconnection phase.
The current sheet half-thickness (Figure 3a) remains approximately $0.3d_{i0}$ and is largely insensitive to the ionization fraction $\chi_0$. This value is thinner than the approximate value of $~0.8d_{i0}$ typically reported in fully kinetic PIC simulations , possibly due to the five-moment model's scalar pressure approximation lacking the full viscous dissipation provided by the non-gyrotropic pressure tensor.
The current sheet length (Figure 3b) shows a weak dependence on $\chi_0$, which diminishes as the system becomes less collisional (increasing $\lambda_{\rm mfp,in}/d_{i0}$). The lengths roughly range between $2d_{i0}$ and $10d_{i0}$, largely consistent with kinetic simulations by~\cite{JJAPOP}.
{Most collisionality sets show a loose tendency toward shorter current sheets at larger \(\chi_0\), but the \(\lambda_{\rm mfp,in}/d_{i0}=5\) set again differs, with the shortest lengths occurring for \(\chi_0=0.02\) and \(0.04\). This is consistent with Figures~\ref{fig:overview} and~\ref{fig:rate_and_onset_thickness_scaling}, where these same cases exhibit more dynamic and faster reconnection, plausibly because additional plasmoid formation breaks the layer into shorter current sheets.}
{These correlations reinforce that plasmoid formation is not only a source of scatter in measured peak rates and lengths, but also an important part of the reconnection dynamics in partially ionized plasmas.}
These results contrast with previous multi-fluid simulations (e.g., \citep{JJAPOP}), which observed a scaling where in-situ current sheet half-thickness and lengths, $\delta_{\rm CS}$ and $L_{\rm CS}$, decrease with ionization like $\chi^{-1/2}$ and $\chi^{-3/4}$ respectively. The five-moment model appears to capture kinetic-like constraints on the current sheet aspect ratio that are absent in standard fluid models.

\begin{figure}
\includegraphics[width=1\columnwidth]{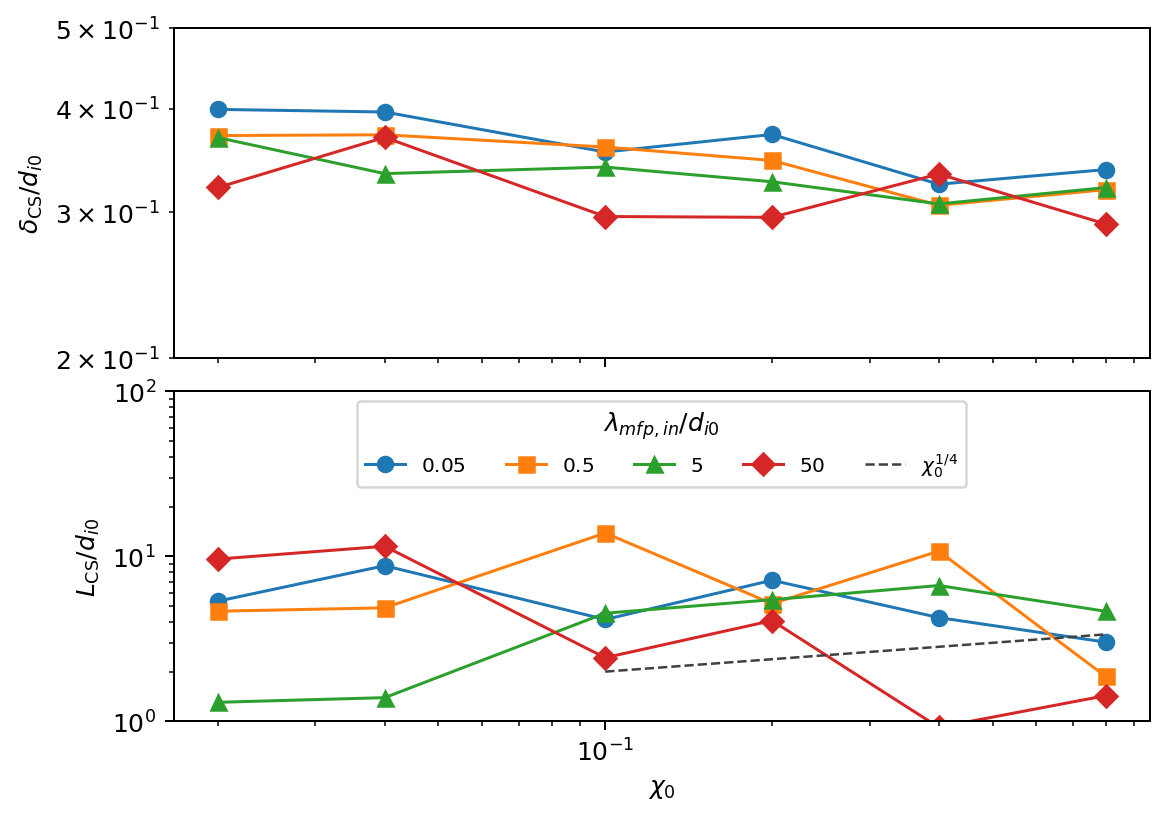}

\caption{Dependence of current sheet morphology on ionization fraction and
collisionality at the time of peak reconnection rate. (a) Current
sheet half-thickness, $\delta_{\rm CS}$, and (b) half-length, $L_{\rm CS}$, plotted
against the initial ionization fraction $\chi_{0}$. Dimensions are
measured at the first primary peak of the reconnection rate to minimize
effects from secondary island formation. The dashed line in panel (b) represents
a $\chi_{0}^{1/4}$ scaling.}\label{fig:thickness_and_length_scaling}
\end{figure}

\begin{figure*}
\includegraphics[width=1\textwidth]{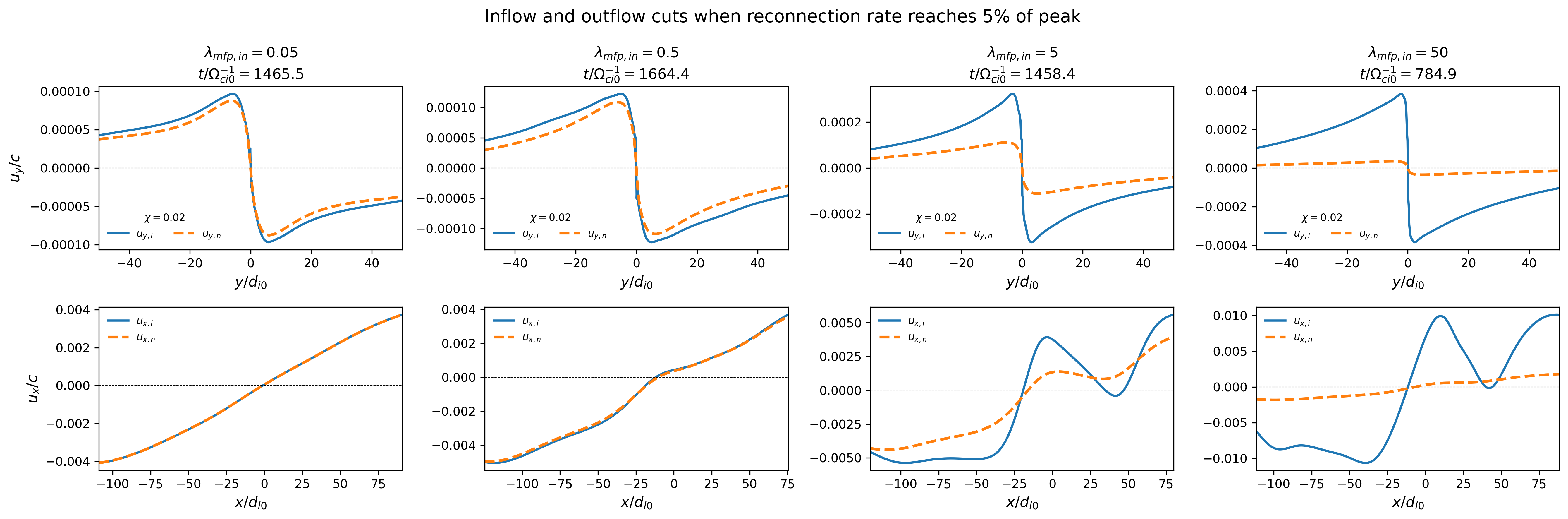}

\includegraphics[width=1\textwidth]{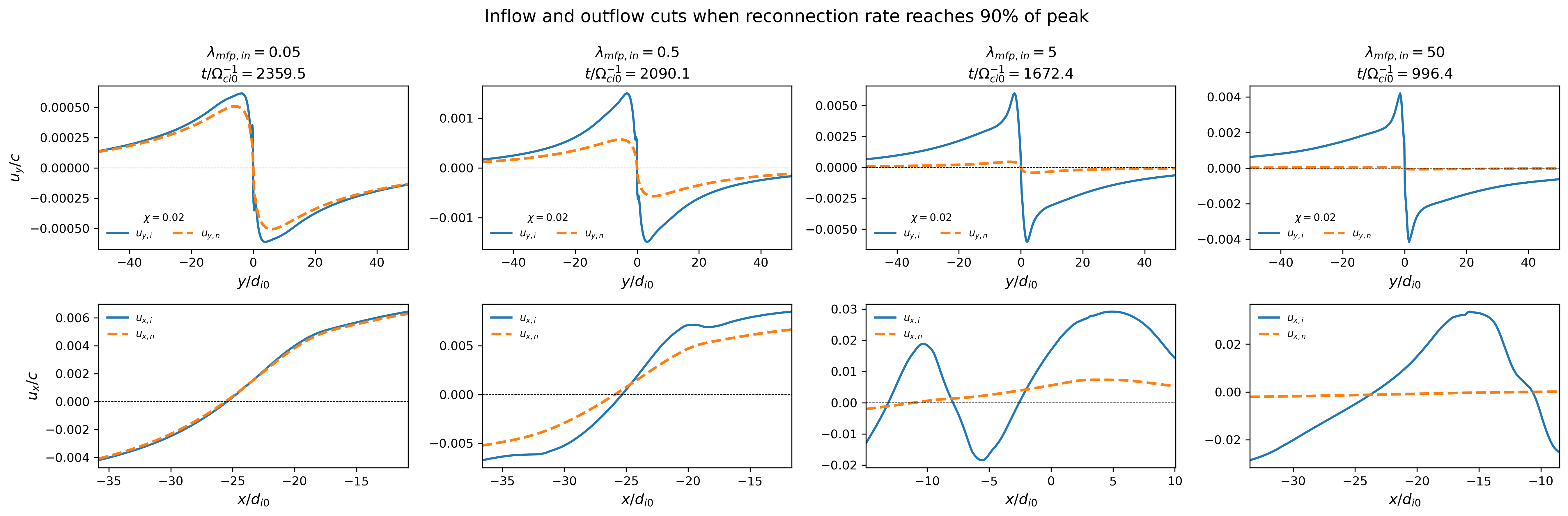}

\caption{Spatial profiles of ion (blue solid) and neutral (orange dashed)
inflow (first and third rows) and outflow (second and bottom rows) velocities in the vicinity of the X-point. (Top
Two Rows) Velocity profiles taken at the pre-onset phase (5\% of peak
reconnection rate). (Bottom Two Rows) Velocity profiles taken during
the fast reconnection phase (90\% of peak rate). All panels display
results for the low ionization case ($\chi_{0}=0.02$), with ion-neutral
collisionality decreasing from left to right ($\lambda_{{\rm mfp,in}}/d_{i0}$
increasing).}\label{fig:inflow_outflow_cuts}
\end{figure*}

Figure \ref{fig:inflow_outflow_cuts} shows inflow and outflow cuts of
ion/neutral velocities for runs with $\chi_{0}=0.02$ (the most weakly
ionized case), with ion-neutral collisionality decreasing from left
to right. 
In the pre-onset phase (top two rows), the inflow region shows clear
ion-neutral decoupling, a signature of ambipolar diffusion that promotes
current-sheet thinning. In the lowest-collisionality run (right),
the species are fully decoupled. Outflow coupling also depends strongly
on collisionality, ranging from fully coupled (left) to decoupled
(right), with the intermediate case ($\lambda_{{\rm mfp,in}}/d_{i0}=0.5$)
showing strong coupling. In the peak reconnection phase (bottom two
rows), accelerating reconnection enhances decoupling in both inflow
and outflow regions. Notably, the intermediate-collisionality case
transitions from coupled to decoupled outflows. Overall, across the
explored ionization fractions and ion-neutral collisionalities, the
three-fluid simulations exhibit a range of outflow coupling behaviors.

\begin{figure*}
\includegraphics[width=1\textwidth]{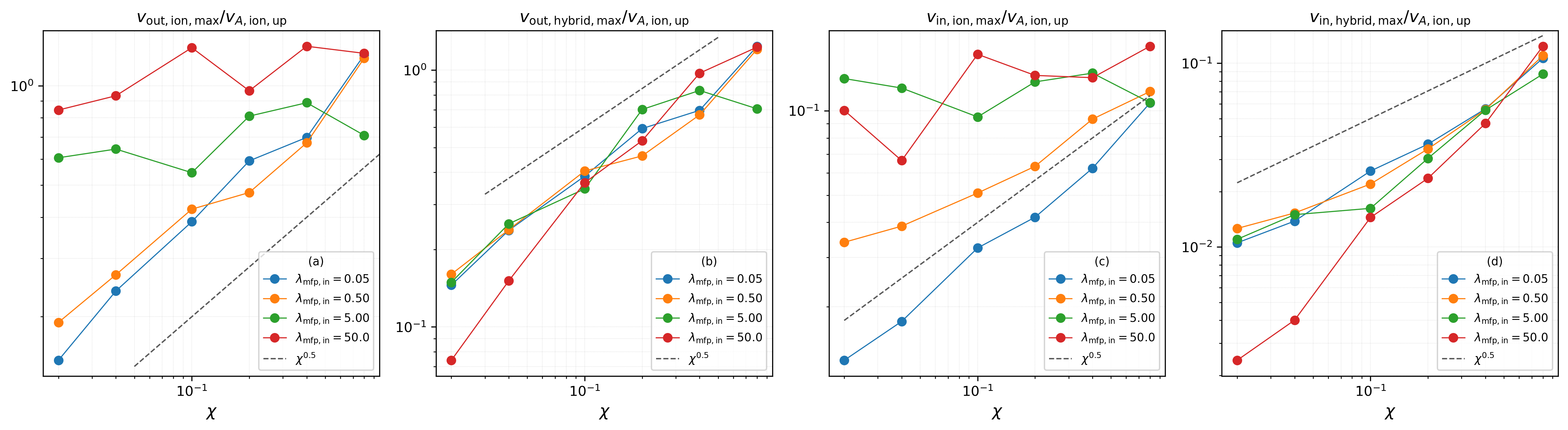}

\caption{Scaling of maximum ion-only and hybrid-ion-neutral inflow/outflow velocities
versus ionization fractions. Different lines correspond to run sets
with different ion-neutral collisionality. The velocities are evaluated
at the time of the first primary peak in each run, measured along
inflow and outflow cuts through the primary X-point. All velocities
are normalized to the initial asymptotic ion-only Alfvén speeds, $v_{A,{\rm ion,up}}$.
Dashed lines indicate the reference $\chi^{1/2}$ scaling.}\label{fig:velocity-scaling}

\end{figure*}

To further understand how ionization fraction influences the velocity
coupling and physical regimes, Figure \ref{fig:velocity-scaling} shows
the scaling of maximum ion
(and hybrid) inflow (and outflow) velocities as functions of $\chi$
for different ion-neutral collisionality.
Here, the hybrid velocity is defined as the mass-weighted bulk velocity
of ions and neutrals, $\mathbf{u}_{{\rm hybrid}}=\left(\rho_{i}\mathbf{u}_{i}+\rho_{n}\mathbf{u}_{n}\right)/\left(\rho_{i}+\rho_{n}\right)$.
For the most strongly coupled
run set with $\lambda_{{\rm mpf,in}}/d_{i0}=0.05$ (blue curve in
panel a), the ion outflow velocity scales like $\chi^{1/2}$ when
normalized by $v_{A,{\rm ion,up}}$, and is approximately 0.9-1 if
normalized by $v_{A,{\rm hybrid,up}}$ (not shown). In the most weakly
coupled runs with $\lambda_{{\rm mpf,in}}/d_{i0}=50$ (red curve in
panel b), the ion outflow velocity is about $0.9v_{A,{\rm ion,up}}$.
In both limits, the ion outflow speed is therefore Alfvénic when measured
relative to the appropriate effective Alfvén speed.

Figure \ref{fig:velocity-scaling}b further shows that, for all collisionality sets, the hybrid
outflow velocity scales approximately as $\chi^{1/2}$ when normalized
by $v_{A,{\rm ion,up}}$. In the strongly coupled limit, this behavior
is consistent with the co-motion of ions and neutrals. As the coupling
gets weaker, the $\chi^{1/2}$ scaling largely persists, with larger
deviations appearing only in the most weakly coupled run set (red
curve). This indicates that the hybrid outflow velocity remains nearly
independent of $\chi$ when normalized by $v_{A,{\rm hybrid,up}}$
over a broad range of collisionality.

Figure \ref{fig:velocity-scaling}c and d show that the ion inflow velocity, when normalized
by $v_{A,{\rm ion,up}}$, shows trends qualitatively similar to those
of the outflow velocities in panels a and b. However, the inflow speeds
display larger departures from the $\chi^{1/2}$ scaling, likely reflecting
the stronger and persistent ion-neutral decoupling in the inflow region.

\section{\label{sec:discussion}Discussion and Conclusion}

In this study, we employed a three-fluid five-moment model with separately evolved electrons, ions, and neutrals to investigate magnetic reconnection in partially ionized plasmas. By simultaneously scanning both the ion-neutral collisionality and ionization fraction, we examined the transition from strongly coupled, resistive reconnection to decoupled, Hall-mediated reconnection, and geometry constraints previously identified in kinetic simulations.

Unlike MHD-based models, the five-moment approach retains independent electron dynamics and associated effects beyond MHD\citep{wang2015comparison,ng2015island}. Our results show that in the weak coupling limit (large $\lambda_{\rm mfp, in}$), the five-moment fluid model reproduces several key kinetic features observed in PIC simulations~\citep{JJAPRL,JJAPOP}, while recovering fluid-theory predictions in well-coupled regimes.

A key finding of this work is the transition of the current sheet structure across coupling regimes. Analytic fluid theories \citep{zweibel2011,Malyshkin2011} and multi-fluid MHD simulations \citep{leake2013, ni2018} often predict that the ion diffusion region broadens to the hybrid scale $d_i \chi^{-1/2}$ in the coupled regime, when ions and neutrals are well coupled. {In contrast, in the weakly coupled and fast-reconnection regimes, our five-moment simulations show that the current sheet instead tends to approach the ion-only inertial length $d_i$, largely independent of ionization fraction.} This finding is consistent with experimental results from MRX \citep{Lawrence2013} and fully kinetic simulations~\citep{JJAPRL,JJAPOP}, suggesting that electron dynamics and other non-MHD effects could allow a fluid model to more faithfully capture the onset of fast reconnection.

These results help bridge the gap between computationally expensive kinetic approaches and large-scale fluid models. We demonstrate that the five-moment formulation can capture the transition to fast reconnection in partially ionized plasmas, a regime of importance in environments such as the solar chromosphere and protostellar disks.
{The plasmoid-coalescence studies of \citet{Murtas2021,Murtas2022}
provide complementary context within the hierarchy of partially ionized plasma
models. Those studies compare fully ionized single-fluid MHD with two-fluid
ion-neutral models, including ionization and recombination in the latter work.
The present study moves in a different direction by using a three-fluid
five-moment model, with separately evolved electrons, ions, and neutrals, in a
Harris-sheet reconnection configuration. These differences in configuration and
model assumptions mean that the results should not be treated as one-to-one
benchmarks, but together they illustrate how reconnection changes as additional
partially ionized plasma physics is retained.}
Future work could extend to a ten-moment model with tensorial pressure for each species. Electron pressure anisotropy and non-gyrotropy are known to influence reconnection in fully ionized plasmas~\citep{Ji2022}, and viscous thermalization is expected to become important in the partially ionized regime by modifying the current-layer structure~\citep{JJAPOP}.
Another important extension is the inclusion of charge exchange, ionization, and recombination~\citep{Murtas2022}, which can dynamically regulate ion-neutral coupling, by, for example, acting as an effective plasma sink that facilitates inflow, thereby reshaping the current sheet and affecting the conditions required for fast reconnection.

Finally, it is important to note that earlier MRX experiments were thought to be limited by the device’s small physical size, with restricted access to the decoupled regime ($\nu_{ni}L/v_A^*<1$). The upgraded and larger Facility for Laboratory Reconnection Experiments (FLARE) \citep{Ji2022,Ji2011} can achieve higher Lundquist numbers and access ion–neutral coupling regimes that were previously out of reach. This presents a valuable opportunity for future laboratory tests of both fluid (MHD and moment-based) and kinetic predictions.

\begin{acknowledgments}
{The open-source code Gkeyll is publicly available at \url{https://github.com/ammarhakim/gkeyll}.} This work was partially supported by NSF grant  AGS-2438328, AGS-2301337, AGS-2301338, DOE grant DE-SC0024639, NASA grant 80NSSC25K0050, the Alfred P. Sloan Research Fellowship, and the IBM Einstein Fellow Fund at the Institute for Advanced Study, Princeton. We would like to acknowledge high-performance computing support from the Derecho system (\url{doi:10.5065/qx9a-pg09}) provided by the NSF National Center for Atmospheric Research (NCAR), sponsored by the National Science Foundation. We also would like to acknowledge high-performance computing support from National Energy Research Scientific Computing Center, a DOE Office of Science user facility, and the NASA High-End Computing (HEC) Program through the NASA Advanced Supercomputing (NAS) Division at Ames Research Center.
\end{acknowledgments}

\bibliography{aipsamp}{}
\bibliographystyle{aasjournalv7}

\end{document}